\documentclass[12pt,superscriptaddress,aps,prd,preprint]{revtex4}
\usepackage{amsmath}
\usepackage{amssymb}
\makeatletter
\usepackage{amsmath}
\usepackage{amssymb}
\usepackage{graphicx}

\newcommand{\bea}{\begin{eqnarray}}
\newcommand{\eea}{\end{eqnarray}}

\newcommand{\pa}{\partial}

\begin{document}
\title{G\"{o}del solution in the bumblebee gravity}

\author{J. R. Nascimento}
\email[]{jroberto@fisica.ufpb.br}
\affiliation{Departamento de F\'{\i}sica, Universidade Federal da 
Para\'{\i}ba\\
 Caixa Postal 5008, 58051-970, Jo\~ao Pessoa, Para\'{\i}ba, Brazil}

\author{A. Yu. Petrov}
\email[]{petrov@fisica.ufpb.br}
\affiliation{Departamento de F\'{\i}sica, Universidade Federal da 
Para\'{\i}ba\\
 Caixa Postal 5008, 58051-970, Jo\~ao Pessoa, Para\'{\i}ba, Brazil}

\author{A. F. Santos}\email[]{alesandroferreira@fisica.ufmt.br}
\affiliation{Instituto de F\'{\i}sica, Universidade Federal de Mato Grosso,\\
78060-900, Cuiab\'{a}, Mato Grosso, Brazil}

\author{W. D. R. Jesus}\email[]{willian.xb@gmail.com}
\affiliation{Instituto de F\'{\i}sica, Universidade Federal de Mato Grosso,\\
78060-900, Cuiab\'{a}, Mato Grosso, Brazil}

\begin{abstract}
Here, we consider a gravity theory involving a spontaneous Lorentz symmetry breaking called the bumblebee model. We show that, at certain values of the bumblebee field, the G\"{o}del metric is consistent within this theory.
\end{abstract}

\maketitle
The general relativity and the standard model are certainly the cornerstones of the modern physics verified through many experimental tests with a very high level of exactness. Despite the success of these theories, there are some open questions, such as, for example, a consistent quantum description of the gravity which has not been found up to now (all known gravity models either non-renormalizable or involve ghosts), modified gravity models which can explain the accelerated expansion of the Universe, and the problem of the unified theory of the fundamental interactions. Within the attempts to solve these problems, the idea of the Lorentz symmetry breaking began to attract the attention. It resulted in formulating the Lorentz-breaking extension of the standard model \cite{Kostelecky1} which involves all known interactions and the Lorentz-violating terms. An excellent review on the extension of the standard model is given in \cite{Bluhm}.

The Lorentz symmetry breaking can be implemented in two different forms: in an explicit form, when the corresponding term presents in the Lagrangian from the very beginning, or in a spontaneous form, when one of vector (in general, tensor) fields acquires a non-zero vacuum expectation value. The spontaneous symmetry breaking methodology is treated as a more elegant form to study Lorentz symmetry violation \cite{Kostelecky2} since in this case the constant vector (or tensor) emerges in a more natural way. The typical model used to study the spontaneous Lorentz symmetry breaking is the bumblebee model discussed in many details in  \cite{Kostelecky3, Bluhm1, Bluhm2,Bertolami}.

By its essence, the bumblebee model involves a vector field $B_\mu$ which acquires a non-zero vacuum expectation value:
\bea
\langle B_\mu \rangle=b_\mu,
\eea
where $b_\mu$ is a constant vector, thus, we have a spontaneous Lorentz symmetry breaking.

The general form of the Lagrangian for the bumblebee model is, see f.e. \cite{Kostelecky3}:
\bea
{\cal L}={\cal L}_0-V(B_\mu B^\mu\mp b^2)+{\cal L}_m,
\eea
where the choice of the sign $\mp$ is discussed further, and the ${\cal L}_0$ is a kinetic term looking like
\bea
{\cal L}_0={\cal L}_{EH}+{\cal L}_{B},
\eea
with ${\cal L}_{EH}$ is an Einstein-Hilbert Lagrangian given by
\bea
{\cal L}_{EH}=\frac{1}{16\pi G}\left[R-2\Lambda\right],
\eea
and ${\cal L}_{B}$ is a Lagrangian for the bumblebee field $B_\mu$ \cite{Will}
\bea
{\cal L}_{B}=\left[\sigma_1B^\mu B^\nu R_{\mu\nu}+\sigma_2B^\mu B_\mu R-\frac{1}{4}\tau_1 B_{\mu\nu}B^{\mu\nu}+\frac{1}{2}\tau_2D_\mu B_\nu D^\mu B^\nu +\frac{1}{2}\tau_3 D_\mu B^\mu D_\nu B^\nu\right],
\eea
with $B_{\mu\nu}=D_\mu B_\nu-D_\nu B_\mu$ is a stress tensor.

Here, we study the Kostelecky-Samuel model \cite{Kostelecky2}, where the Lagrangian ${\cal L}_0$ assumes a form of the Einstein-Maxwell Lagrangian which reduces to the Maxwell Lagrangian in the flat space, so, the parameters $\sigma_1$, $\sigma_2$, $\tau_2$ e $\tau_3$ vanish, while $\tau_1=1$. So, we rest with
\bea
{\cal L}_0^{KS}\equiv{\cal L}_{KS}=\frac{1}{16\pi G}(R-2\Lambda)-\frac{1}{4} B_{\mu\nu}B^{\mu\nu}.
\eea
The spontaneous Lorentz symmetry breaking is generally induced through a potential in the corresponding Lagrangian \cite{Bluhm3}. There are different possibilities for choice of the potential, here we choose the quartic one:
\bea
V=\frac{1}{2}\kappa\left(B_\mu B^\mu\mp b^2\right)^2,
\eea
where $\kappa$ and $b^2$ are essentially positive constants to provide the existence of minima. This potential, first, does not involve any extra field, second, vanishes at the minimum together with its first derivative.  The $\mp$ sign reflects the fact that the vector $B^{\mu}$ can be time-like as well as space-like (in the G\"{o}del case we study here, time-like $B_{\mu}$ yields $B^{\mu}B_{\mu}<0$ since for the time-like $B_{\mu}=(b,0,0,0)$, and $g^{00}=-\frac{1}{a^2}$, one has $B^{\mu}B_{\mu}<0$). In this case, our Lagrangian takes the form
\bea
{\cal L}=\frac{1}{16\pi G}(R-2\Lambda)-\frac{1}{4} B_{\mu\nu}B^{\mu\nu}+\frac{1}{2}\kappa\left(B_\mu B^\mu\pm b^2\right)^2+{\cal L}_m.
\eea

The dynamics of the gravitational sector is described by the Einstein equations:
\bea
G_{\mu\nu}=8\pi G T_{\mu\nu},\label{mov}
\eea
where $G_{\mu\nu}$ is an usual Einstein tensor
\bea
G_{\mu\nu}=R_{\mu\nu}-\frac{1}{2}g_{\mu\nu}R,\label{einstein}
\eea
and $T_{\mu\nu}$ is an energy-momentum tensor of the matter composed by two parts:
\bea
T_{\mu\nu}=T_{\mu\nu}^M+{\cal T}_{\mu\nu}^B,
\eea
where $T_{\mu\nu}^M$ is an energy-momentum tensor of the usual matter (for example, relativistic fluid), and ${\cal T}_{\mu\nu}^B\equiv \frac{T_{\mu\nu}^B}{8\pi G}$ corresponds to the bumblebee field, so that $T_{\mu\nu}^B$ is given by 
\cite{Kostelecky3,Seifert}:
\bea
T_{\mu\nu}^B=B_{\mu\alpha}B^{\phantom{\nu}\alpha}_\nu-\frac{1}{4}g_{\mu\nu}B_{\alpha\beta}B^{\alpha\beta}-Vg_{\mu\nu}+2V'B_\mu B_\nu, \label{tensorB}
\eea
where $V'$ is a derivative of the potential with respect to the argument, that is, $V'=\frac{\partial V}{\partial X}\Bigl|_{X=B_\mu B^\mu\pm b^2}$.

In this paper, we verify the consistency of the G\"{o}del metric \cite{Godel} within the bumblebee gravity. The fundamental feature of this solution is the presence of the closed timelike curves (CTCs) which inspired an intensive discussions on the causality and time travels in the gravity context(on these issues see f.e. \cite{timetravel}). Besides of this, the  G\"{o}del solution is interesting since it describes an Universe with the rotating matter and non-zero cosmological constant. An extensive discussion of causality aspects of the G\"{o}del metric can be found in \cite{Reb}. Some issues related to consistency of the G\"{o}del solutions in different gravity models are discussed in \cite{ourGodel}.

The G\"{o}del metric looks like
\bea
ds^2=a^2\Bigl[dt^2-dx^2+\frac{1}{2}e^{2x}dy^2-dz^2+2 e^x dt\,dy\Bigl],\label{godel}
\eea
where $a$ is a constant. To study the Einstein equation (\ref{mov}) for a metric (\ref{godel}), we need to find the non-zero components of the Einstein tensor (\ref{einstein}). To do it, we write down the non-zero components of the Ricci tensor:
\bea
R_{00}&=&1\nonumber\\
R_{02}&=&R_{20}=e^x\nonumber\\
R_{22}&=&e^{2x},\label{Ricci}
\eea
with the scalar curvature is
\bea
R=\frac{1}{a^2}.\label{scalar}
\eea
With (\ref{Ricci}) and (\ref{scalar}), we determine the left-hand side of the Eq. (\ref{mov}), that is, the geometrical part. It remains to find the energy-momentum tensor of the matter. One of its parts is the energy-momentum tensor of the relativistic fluid which was introduced in the original paper \cite{Godel}: 
\bea
 T_{\mu\nu}^M=\rho u_\mu u_\nu+\Lambda' g_{\mu\nu},\label{matter}
\eea
where $u_\mu=(a, 0, ae^x, 0)$ e $\Lambda'=\frac{\Lambda}{8\pi G}$, with $\Lambda$ is a cosmological constant. Another part of the energy-momentum tensor is associated to the bumblebee field $B_\mu$ given in (\ref{tensorB}). To find the components of the energy-momentum tensor, we choose the form of the field $B_\mu$, which, in the vacuum, must satisfy
\bea
g^{\mu\nu}B_\mu B_\nu=\pm b^2, \label{condition}
\eea
to provide a minimum for the potential.

As a first attempt, we choose
\bea
B_\mu=(ab, 0, 0, 0),\label{first}
\eea
which evidently satisfies a condition of the vacuum (\ref{condition}). Therefore, the potential vanishes, and the energy-momentum tensor takes the form
\bea
T_{\mu\nu}^B=B_{\mu\alpha}B_\nu\,^{\alpha}-\frac{1}{4}g_{\mu\nu}B_{\alpha\beta}B^{\alpha\beta}.\label{newtensor}
\eea
With a choice (\ref{first}) we have that only $B_0\neq 0$ is a constant, thus, the stress tensor $B_{\mu\nu}=\partial_\mu B_\nu-\partial_\nu B_\mu=0$ vanishes (we remind that $\nabla_{\mu}B_{\nu}-\nabla_{\nu}B_{\mu}=\partial_{\mu}B_{\nu}-\partial_{\nu}B_{\mu}$, with the terms with the Christoffel symbols cancel each other), so, one has
\bea
T_{\mu\nu}^B=0,
\eea
i.e., all components of the energy-momentum tensor  of the field $B_\mu$ are zero. 
We note that in principle the bumblebee field must satisfy also its equations of motion \cite{Seifert}:
\bea
\nabla_{\mu}B^{\mu\nu}=2V^{\prime}(B^2)B^{\nu},
\eea
however, for the constant $B^{\mu}$ corresponding to the vacuum, and for our choice of the potential, this equation is identically satisfied.

The situation does not differ, that is, we again have the fields $B_{\mu}$ corresponding to the vacuum, together with vanishing of the energy-momentum tensor, $T_{\mu\nu}^B=0$ because of the annihilation of the tensor $B^{\mu\nu}$ equation of motion for $B^{\mu}$ identically satisfied, for two other choices of $B_{\mu}$:
\bea
B_\mu&=&(0, ab, 0, 0),\label{second}\\
B_\mu&=&(0, 0, 0, ab).\label{third}
\eea
Therefore, for these three choices of the $B_\mu$ field, (\ref{first}), (\ref{second}) and (\ref{third}), the equations of motion  (\ref{mov}) are reduced to those ones obtained by G\"{o}del in \cite{Godel}, thus, we conclude that in these cases the  G\"{o}del metric is consistent with the spontaneous Lorentz breaking. So, the CTCs persist in this case in the same way as in  \cite{Godel}.
Thus, we convinced that all values of the $B_{\mu}$ field corresponding to the vacuum and vanishing of the $B_{\mu\nu}$ tensor at the same time are compatible with the G\"{o}del metric. Indeed, in these cases the equations of motion of the bumblebee field are satisfied identically, and its energy-momentum tensor vanishes, thus, the additive terms in the Einstein equations involving the bumblebee field completely disappear.

Now, let us try the more sophisticated situation, that is, let us suppose that the $B_{\mu}$ field corresponds to the vacuum but the $B_{\mu\nu}$ does not vanish (in particular, this is the case when $B_{\mu}$ is directed along $y$ axis). We note that if we restrict our consideration to the vacuum case (otherwise we will have no spontaneous Lorentz symmetry breaking but only a simple dynamics of the bumblebee field away from the vacuum which does not principally differ from other fields, and this situation is not interesting for us since our aim consist namely in keeping track of the spontaneous Lorentz symmetry breaking), with the quartic potential as it has been done above, our equation of motion reduces to
\bea
\partial_{\mu}B^{\mu\nu}+\Gamma^{\mu}_{\mu\lambda}B^{\lambda\nu}=0.
\eea
It is known that for the G\"{o}del metric, the only non-zero Christoffel symbol with coinciding upper and lower indices is $\Gamma^0_{01}=1$. So, if we suggest all fields to depend only on $x_1=x$ as the components of the G\"{o}del metric do, we get
\bea
\partial_1B^{1\nu}+B^{1\nu}=0,
\eea
thus, the solution for the stress tensor is $B^{1\nu}=k^{\nu}e^{-x}$, with $k^{\nu}$ is a some constant vector. It is clear that its $x$ component vanishes, $k^1=0$, which means that $B^1$ component cannot be restricted from this equation. Actually, we will determine appropriate restrictions on $B^1$ afterwards. Further, since there is no dependence on other coordinates, the value of $B^{\mu}$ corresponding to this stress tensor yields $B^{1\nu}=\pa^1 B^{\nu}=k^{\nu}e^{-x}$, so, one has $B^{\nu}=-k^{\nu}e^{-x}$ (with, again, $\nu\neq 1$).

Now, it remains to find whether such a $B^{\nu}$ can yield a vacuum condition (\ref{condition}). Taking into account the structure of the G\"{o}del metric, one finds that the (\ref{condition}) implies in the following relation for $B^{\mu}$ 
\bea
a^2((B^0)^2-(B^1)^2+2e^xB^0B^2+\frac{1}{2}e^{2x}(B^2)^2-(B^3)^2)=\pm b^2.
\eea
If we take into account that $B^{\nu}=-k^{\nu}e^{-x}$ (except of $\nu=1$), with a constant $k^{\nu}$, one gets from this:
\bea
a^2((k^0)^2e^{-2x}-(B^1)^2+2e^{-x}k^0k^2+\frac{1}{2}(k^2)^2-e^{-2x}(k^3)^2)=\pm b^2.
\eea
It implies $B^1\equiv-b_1=const$, $k^0k^2=0$ and $(k^0)^2-(k^3)^2=0$. We denote $k^2\equiv q$ (one should remind that it is a constant) The simplest choice is $k^0=k^3=0$, i.e. $B^0=B^3=0$, with the vacuum condition $g_{\mu\nu}B^{\mu}B^{\nu}=\pm b^2$ takes the form
\bea
\label{cond}
a^2(\frac{q^2}{2}-(b_1)^2)=\pm b^2.
\eea 
We note that, if we try other choice of the vacuum, that is, $k^0=k^3=\frac{d}{a}$, with $k^2=0$, we  will have $B_{01}=-B_{31}=-ade^{-x}$, which gives $T_{33}^B=0$, but  $T^B_{11}=-d^2e^{-2x}$, $T^B_{00}=-2d^2e^{-2x}$, $T^B_{02}=-d^2e^{-x}$, $T^B_{22}=-\frac{d^2}{2}$. Taking into account the explicit form of the Einstein tensor (in our case, its components involve the same exponentials as the correspondent components of the Ricci tensor and the metric tensor), we see that such a form of the energy-momentum tensor of the bumblebee field is inconsistent with the Einstein equations in the G\"{o}del metric case while $d\neq 0$.

So, we restricted the vacuum field $B^{\mu}$ to be $B^{\mu}=(0,-b_1,-q e^{-x},0)$, with $q=\pm\sqrt{2b^2_1\pm\frac{2b^2}{a^2}}$. It is more convenient to use the lower-index vector $B_{\mu}=(-a^2q,a^2b_1,-\frac{a^2q}{2}e^x,0)$, with $q$ and $b_1$ are related through (\ref{cond}).
Let us verify the consistency of this case with the Einstein equations. We get the only non-zero component of the stress tensor $B_{12}=-B_{21}=-\frac{a^2q}{2}e^x$. So, the energy-momentum tensor of the bumblebee field is reduced to (\ref{newtensor}), whose non-zero components have the following explicit form
\bea
T_{00}^B&=&-\frac{a^2q^2}{4},\quad\;
T_{02}^B=-\frac{a^2q^2}{4}e^x,\nonumber\\
T_{11}^B&=&-\frac{a^2q^2}{4},\nonumber\\
T_{22}^B&=&-\frac{3}{8}a^2q^2e^{2x},\quad\;
T_{33}^B=\frac{a^2q^2}{4}.\label{B2}
\eea
And the non-zero components of the ``usual'' matter energy-momentum tensor (\ref{matter}) are:
\bea
T_{00}^M&=&\rho a^2+\Lambda'a^2,\quad\;
T_{02}^M=\rho a^2e^x+\Lambda'a^2e^x,\nonumber\\
T_{11}^M&=&-\Lambda'a^2,\nonumber\\
T_{22}^M&=&\rho a^2e^{2x}+\frac{1}{2}\Lambda'a^2e^{2x},\quad\;
T_{33}^M=-\Lambda'a^2.\label{M2}
\eea
Using Eqs. (\ref{Ricci}), (\ref{scalar}), (\ref{B2}) and (\ref{M2}), we get the following modified Einstein equations:
\bea
(00):\, \frac{1}{2}&=&8\pi G\rho a^2+\Lambda a^2-\frac{a^2q^2}{4},\\
(11):\,\frac{1}{2}&=&-\Lambda a^2-\frac{a^2q^2}{4},\nonumber\\
(22):\,\frac{3}{2}&=&16\pi G\rho a^2+\Lambda a^2-\frac{3}{8}a^2q^2,\nonumber\\
(33):\,\frac{1}{2}&=&-\Lambda a^2+\frac{a^2q^2}{4}.\label{33}\nonumber
\eea
The equation for the (02) component is identical to that one for the (00) component with only difference in the overall factor $e^x$ both in left-hand side and right-hand side of the equation.

We see that the system is overdetermined (it involves only three variables but four equations) -- it is interesting to note that actually, the system of equation in the case of the G\"{o}del metric is overdetermined also in other gravity models, such as Chern-Simons gravity and the Horava-Lifshitz gravity \cite{ourGodel}. The only consistent solution corresponds to the case $q=0$ when the system is reduced to the usual case \cite{Godel}, and so, $8\pi G\rho=\frac{1}{a^2}$, $\Lambda=-4\pi G\rho$. 
It is interesting to discuss now the case $q=0$. The situation with a nontrivial symmetry breaking (that is, $b\neq 0$) is when the potential is $V=\frac{\kappa}{2}(B^{\mu}B_{\mu}+b^2)^2$, so, $B^{\mu}=(0,-b_1,0,0)$ is space-like, and the sign in the r.h.s. of (\ref{cond}) is $(-)$, i.e. one has $b^2_1=\frac{b^2}{a^2}$. It means that in this case the bumblebee field is directed just along the $x$ axis, and, moreover, it is constant, so, indeed, the stress tensor for $B_{\mu}$ is forced to vanish. Actually, we have showed that the G\"{o}del solution is consistent with the bumblebee gravity only if the stress tensor vanishes.

Let us make the conclusions. We have shown that in the bumblebee gravity, the G\"{o}del solution can be found for one of the vacua of the theory only if the stress tensor of the bumblebee field vanishes. Therefore, we found that the spontaneous breaking of the Lorentz symmetry can be compatible with the existence of the CTCs only in certain situations, that is, if $B_{\mu}$ is a constant vector (so, the stress tensor $B_{\mu\nu}$ vanishes) corresponding to the vacuum, while in other cases the consistency is jeopardized. Therefore, we can conclude that some forms of the spontaneous symmetry breaking allow to rule out the CTCs.

{\bf Acknowledgments.}
This work was partially supported by Conselho Nacional de Desenvolvimento Cient\'{\i}fico e Tecnol\'{o}gico (CNPq). A. Yu. P. has been supported by the CNPq project  303438/2012-6 and A. F. S. has been suported by the CNPq project 476166/2013-6.

\end{document}